\documentclass[aps,floatfix,epsfig]{revtex4}
\usepackage{pdfpages}
\usepackage{subfigure}
\usepackage{amsmath}
\usepackage{eqnarray,amsmath}
\usepackage{amssymb}
\usepackage{amsthm}
\usepackage{epstopdf}
\usepackage{mathrsfs}
\usepackage{natbib}
\usepackage{amssymb,latexsym}
\usepackage{dcolumn}
\usepackage{graphicx}
\usepackage{url}
\usepackage{hyperref}
\usepackage{verbatim}
\usepackage{mathrsfs}
\usepackage{amsfonts}
\usepackage{latexsym}

\def\be{\begin{equation}}
	\def\ee{\end{equation}}
\def\ba{\begin{eqnarray}}
	\def\ea{\end{eqnarray}}

\begin{document}
\title{Searching for the hydrogen 21 cm line in cosmos}
\author{Mahta Moazzenzadeh}
 \affiliation{Department of Physics, K. N. Toosi University of Technology, P.O. Box 15875-4416, Tehran, Iran}
\email{Moazzenzade@kntu.ac.ir}

\author{Javad T. Firouzjaee} 
\email{firouzjaee@kntu.ac.ir}
\affiliation{Department of Physics, K. N. Toosi University of Technology, P.O. Box 15875-4416, Tehran, Iran}
\affiliation{School of Physics, Institute for Research in Fundamental Sciences (IPM), P.O. Box 19395-5531, Tehran, Iran}

\begin{abstract}
\textbf{Abstract:} It is widely agreed that studying the 21 cm emission line from neutral hydrogen may be our best hope for understanding the creation of the first structures during the dark ages and the Epoch of Reionization (EoR). This hyperfine transition occurs as a result of the spin-spin interaction between the electron and the proton in hydrogen: The energy of the parallel spins (triplet) state is greater than that of the anti-parallel spins (singlet). This transition is strongly banned, with a transition probability of $2.6 \times 10^{-15} \mathrm{s}^{-1}$ being exceedingly low. Despite its low chance, the 21 cm hyperfine transition is one of the most important tools in observational astronomy due to the abundance of hydrogen in the Universe. The relative occupation number of the ground state and the occupation number of the excited state determine the strength of the 21 cm emission or absorption.  Collisions or the so-called Wouthuysen-Field effect can both excite the 21 cm line.
 In this article, we give a brief review of the basic physics of  21 cm emission and highlight the current main worldwide 21 cm signal experiments. We also discuss  several  related concepts in astronomy and cosmology with these observations.
\end{abstract}
%
%
\maketitle
\tableofcontents
\section{Introduction}
\subsection{A brief introduction on 21 cm emission}

As hydrogen is the main component of interstellar gas, we often classify regions of space according to the amount of neutral or ionized hydrogen in them. Ionized hydrogen clouds are known as H II regions. (Scientists who work with spectra use the Roman numeral I to indicate that an atom is neutral; successively higher Roman numerals are used for each higher stage of ionization. H II signifies hydrogen which has lost one electron.)

During each downward transition between the multiple energies of hydrogen atoms, electrons give up a portion of their energy in the form of light. A process by which ultraviolet light is converted into visible light is called fluorescence. Interstellar gas contains many other elements besides hydrogen. Sometimes, they are also ionized near hot stars; they capture electrons and emit light just like hydrogen, making them observable. But generally, the red hydrogen lines are the strongest, which is why H II regions appear red
\cite{pressbooks}.

Fluorescent lights on Earth use the same mechanism as fluorescent H II regions. When you turn on the current,  electrons collide with atoms of mercury vapor in the tube. The mercury becomes excited to a high-energy state as a result of these collisions. The mercury atoms emit some of their energy as ultraviolet photons when they return to lower-energy levels, which strikes a screen coated with phosphor on the inside wall of the tube. (These atoms emit a wider range of light colors than those in H II, giving fluorescent lights their characteristic white glow. Hydrogen atoms in an H II region have a more limited range of colors.) Only very few of the stars in the interstellar medium are hot enough to produce H II regions; most of the interstellar medium is filled with neutral (non-ionized) hydrogen. In what way should we search for it?

There is no visible light emitted or absorbed by neutral hydrogen atoms at the temperatures of interstellar space. The same holds true for most other trace elements in the interstellar hydrogen mixture. Despite the interstellar temperatures, some of these other elements have the ability to absorb visible light. As a consequence, we may sometimes see additional lines in the spectrum of a bright source such as a hot star or a galaxy when interstellar gas absorbs light at specific frequencies. Insufficient sensitivity, we can also detect many other stellar absorption lines, besides calcium and sodium
\cite{lumenlearning}.

Including the redshift, this line will be observed at frequencies from 200 MHz to about 9 MHz on Earth. Hydrogen lines are a highly valuable tool in Big Bang theory because it allows us to figure out what happened during the "dark ages." The analysis of the incoming redshifted 21 cm radiation could serve two purposes, as well. First, it could provide a very precise account of the matter spectrum in the period following recombination by mapping the intensity of the incoming redshifted 21 cm radiation. Second, it provides evidence of radiation produced by stars or quasars that ionized neutral hydrogen, since these stars and quasars will appear as holes in the 21 cm background.\\
 

\subsection{Prediction of 21 cm line radiation}

As a result of Grote Reber's radio astronomical discoveries, the great Dutch astronomer Jan Oort recognized that a radio spectral line could be used to discover the structure of our galaxy. During the course of his studies, Oort used optical means to study the rotation and structure of the galaxy. He noticed the immense amount of dust accumulating in the galactic plane, which blocked visible light. Due to the absorption of light by distant stars, we can see only a few thousand light years towards the galactic center. However, radio waves can penetrate the dust and show us the galactic center and actually the other side of our galaxy. By the Doppler effect, the frequency of the line shifts, which enables the measurement of the velocity of the gas. Using differential rotation, one can estimate the distance to gas clouds in the galaxy and map the distribution of matter within.

Van de Hulst was given a task by Oort to find out if there were any radio spectral lines and at what frequency they would exist. Since hydrogen is the most abundant element, this was the element he studied first. According to his calculations, a "hyper-fine" transition in the ground state of hydrogen produces radiation at 1420 MHz, generally occurring at a wavelength of about 21 cm. Electrons can either have their magnetic moment parallel to or anti-parallel to those of their proton counterparts in hydrogen's ground state. Because of this, when the electron transitions to the anti-parallel state, it produces 21 cm wavelength radiation \cite{nrao}.

In 1951, Ewen and Purcell discovered the 21 cm line (1420.4 MHz), whose findings then were confirmed by Dutch astronomers Muller and Oort, and by Christiansen and Hindman in Australia. It wasn't until 1952 that the first maps of neutral hydrogen in the Galaxy were made, and the spiral structure of the Milky Way was uncovered. 


The goal of this review is to introduce the main projects to detect the 21 cm signal in cosmology and astrophysics. Search on 21 cm physics continues to be a very active field due to the upcoming telescopes data.
This review is organized as follows: in Section II, we explain the basic physics of the hydrogen 21 cm emission line. In Section III, the main experiments to detect the 21 cm emission line are reviewed. The role of 21 cm emission line in galaxy physics is explained in Section IV. In Section V, we study the 21 cm emitted from the sun. Section VI is devoted to probing the dark matter and dark energy physics in 21 cm emission line. Finally, we conclude and summarize the main results in Section VII.\\

\section{Physics of the 21 cm emission}
\subsection{Basic physics} 

Since hydrogen radiation penetrates the dust clouds at 1420 MHz, we get a more complete picture of hydrogen than the stars themselves, whose visible light cannot penetrate the dust clouds.

A radiation of 1420 MHz is generated from the interaction between the electron spin and the nuclear spin in hydrogen 1S orbital. It represents the transition between the two levels of this ground state. This splitting is known as hyper-fine structure. Hydrogen in its lower state, as a result of its quantum properties, will absorb 1420 MHz, and its emission indicates that a prior excitation occurred in the upper state. 


Because of the mass in the exponent, the proton also has a magnetic dipole moment, though that moment is much smaller than that of the electron which can be written \cite{Hyperfine splitting in the ground state of hydrogen}

\begin{equation}
\boldsymbol{\mu}_{p}=\frac{g_{p} e}{2 m_{p}} \mathbf{S}_{p}, \quad \boldsymbol{\mu}_{e}=-\frac{e}{m_{e}} \mathbf{S}_{e},
\end{equation}

where ${\mu}_{p}$ is the proton magnetic dipole moment and ${\mu}_{e}$ is the electron magnetic dipole moment. Also, ${g}_{p}$ indicates the proton's $g$-factor, ${g}_{e}$ represents the electron's $g$-factor, $\mathbf{S}_{p}$ is the spin of the proton, $\mathbf{S}_{e}$ indicates the electron's spin. ${m_{e}}$ is the mass of the electron and ${m_{p}}$ is the mass of the proton.

(The proton is a composite structure composed of three quarks, and it has a different gyromagnetic ratio than the electron-that is why its $g$-factor, $\left(g_{p}\right)$, is $5.59$, not the electron's $2.00$). As a matter of electrodynamics, a dipole $\mu$ generates a magnetic field:

\begin{equation}
\mathbf{B}=\frac{\mu_{0}}{4 \pi r^{3}}[3(\boldsymbol{\mu} \cdot \hat{r}) \hat{r}-\boldsymbol{\mu}]+\frac{2 \mu_{0}}{3} \boldsymbol{\mu} \delta^{3}(\mathbf{r}) ,
\end{equation}

where ${\mu_{0}}$ is vacuum permeability and ${\mu}$ is Magnetic permeability.
Thus, in the magnetic field created by a proton's dipole moment, the electron would have the following Hamiltonian:

\begin{equation}
H_{\mathrm{hf}}^{\prime}=\frac{\mu_{0} g_{p} e^{2}}{8 \pi m_{p} m_{e}} \frac{\left[3\left(\mathbf{S}_{p} \cdot \hat{r}\right)\left(\mathbf{S}_{e} \cdot \hat{r}\right)-\mathbf{S}_{p} \cdot \mathbf{S}_{e}\right]}{r^{3}}+\frac{\mu_{0} g_{p} e^{2}}{3 m_{p} m_{e}} \mathbf{S}_{p} \cdot \mathbf{S}_{e} \delta^{3}(\mathbf{r}) .
\end{equation}

First-order corrections to energy are defined according to perturbation theory as the expectation value of the perturbing Hamiltonian:

\begin{equation}
\begin{aligned}
E_{\mathrm{hf}}^{1}=& \frac{\mu_{0} g_{p} e^{2}}{8 \pi m_{p} m_{e}}\left\langle\frac{3\left(\mathbf{S}_{p} \cdot \hat{r}\right)\left(\mathbf{S}_{e} \cdot \hat{r}\right)-\mathbf{S}_{p} \cdot \mathbf{S}_{e}}{r^{3}}\right) \\
&+\frac{\mu_{0} g_{p} e^{2}}{3 m_{p} m_{e}}\left\langle\mathbf{S}_{p} \cdot \mathbf{S}_{e}\right\rangle|\psi(0)|^{2} .
\end{aligned}
\end{equation}

As the wave function is spherically symmetric in the ground state (or in any other state that $l=0$ ) the first expectation value disappears. In the meantime, from Equation 

\begin{equation}
\psi_{100}(r, \theta, \phi)=\frac{1}{\sqrt{\pi a^{3}}} e^{-r/a} ,
\end{equation}

we find that $\left|\psi_{100}(0)\right|^{2}=1 /\left(\pi a^{3}\right)$, hence

\begin{equation} \label{Ehr}
E_{\mathrm{hr}}^{1}=\frac{\mu_{0} g_{p} e^{2}}{3 \pi m_{p} m_{e} a^{3}}\left\langle\mathbf{S}_{p} \cdot \mathbf{S}_{e}\right\rangle ,
\end{equation}

in the ground state. The spin-spin coupling involves the dot product of two spins (contrasted with the spin-orbit coupling that involves $\mathbf{S} \cdot \mathbf{L}$ ).

Spin-spin coupling results in no more conservation of individual spin angular momenta; the "good" states are the eigenvectors of the total spin,

\begin{equation}
\mathbf{S} \equiv \mathbf{S}_{e}+\mathbf{S}_{p}.
\end{equation}

As before, we square this out to get

\begin{equation}
\mathbf{S}_{p} \cdot \mathbf{S}_{e}=\frac{1}{2}\left(S^{2}-S_{e}^{2}-S_{p}^{2}\right).
\end{equation}

Both the electron and proton have a spin of $1 / 2$, so $S_{e}^{2}=S_{p}^{2}=(3 / 4) \hbar^{2}$. In the triplet state (spins "parallel") the total spin is $ 1 $. As a result,  $S^{2}=2 \hbar^{2}$; in the singlet state the total spin is 0, and $S^{2}=0$. Therefore, from Eq. \eqref{Ehr} we get

\begin{equation}
E_{\mathrm{hf}}^{1}=\frac{4 g_{p} \hbar^{4}}{3 m_{p} m_{e}^{2} c^{2} a^{4}}\left\{\begin{array}{ll}
+1 / 4, & (\text { triplet }) \\
-3 / 4, & \text { (singlet) }
\end{array}\right.
\end{equation}

Spin-spin coupling eliminates spin degeneracy of the ground state, thereby promoting the triplet and depressing the singlet. Evidently, the energy gap is 

\begin{equation}
\Delta E=\frac{4 g_{p} \hbar^{4}}{3 m_{p} m_{e}^{2} c^{2} a^{4}}=5.88 \times 10^{-6} \mathrm{eV}.
\end{equation}

In the transition from a triplet state to a singlet state, a microwave photon is emitted of a frequency 

\begin{equation}
v=\frac{\Delta E}{h}=1420 \mathrm{MHz}.
\end{equation}

Among all the radiation occurring throughout the universe, this 21 cm line is one of the most ubiquitous.

The line center frequency is:

\begin{equation}
\label{freq}
\nu_{10} = {8 \over 3} g_{\rm I}
\biggl( {m_{\rm e} \over m_{\rm p} }\biggr) \alpha^2 (R_{\rm M} c)
\approx1420.405751{\rm ~MHz},
\end{equation}
where
$g_{\rm I} \approx 5.58569$ is the  nuclear $g$-factor for a proton,
$\alpha \equiv e^2 / (\hbar c) \approx 1 / 137.036$ is the dimensionless fine-structure constant,
and $R_{\rm M}c$ is the hydrogen Rydberg frequency. After simplifying we get, 

\begin{equation} \label{freq1}
\nu_{10}
\approx 1420.4
{\rm ~MHz}.
\end{equation}

The 21 cm line spin temperature $T_{S}$ corresponds to its excitation temperature. It is defined by the ratio between the number densities of hydrogen atoms $n_{i}$ in the two hyper-fine levels (which we label with a subscript 0 and 1 for the $1 S$ singlet and $1 S$ triplet levels, respectively)
\begin{equation}
n_{1} / n_{0}=\left(g_{1} / g_{0}\right) \exp \left(-T_{\star} / T_{S}\right)
\end{equation}

where $\left(g_{1} / g_{0}\right)=3$ is the ratio of the statistical degeneracy factors of the two levels, and $T_{\star} \equiv h c / k \lambda_{21 \mathrm{cm}}=0.068 \mathrm{K}$ With this definition, a cloud of hydrogen has an optical depth of 

\begin{equation}
\tau_{\nu}=\int \mathrm{d} s\left[1-\exp \left(-E_{10} / k_{B} T_{S}\right)\right] \sigma_{0} \phi(\nu) n_{0},
\end{equation}

where $n_{0}=n_{H} / 4$ with $n_{H}$ being the hydrogen density, and we have denoted the $21 \mathrm{cm}$ cross-section as $\sigma(\nu)=\sigma_{0} \phi(\nu),$ with $\sigma_{0} \equiv 3 c^{2} A_{10} / 8 \pi \nu^{2},$ where $A_{10}=2.85 \times 10^{-15} \mathrm{s}^{-1}$
indicates the rate at which the spin-flip transition spontaneously decays, so that $\int \phi(\nu) \mathrm{d} \nu=1 .$ 

The spin temperature can be calculated by the following formula: 

\begin{equation}
T_{S}^{-1}=\frac{T_{\gamma}^{-1}+x_{\alpha} T_{\alpha}^{-1}+x_{c} T_{K}^{-1}}{1+x_{\alpha}+x_{c}}
\end{equation}

where $T_{\gamma}$ is the temperature of the surrounding bath of radio photons, typically determined by the CMB so that $T_{\gamma}=T_{\mathrm{CMB}} ; T_{\alpha}$ is the color temperature of the Lyman-alpha line (typically denoted by Ly-$\alpha$) radiation field at the Ly-$\alpha$ frequency and is closely coupled to the gas kinetic temperature $T_{K}$ by recoil during repeated scattering, and $x_{c}, x_{\alpha}$ are the coupling coefficients due to atomic collisions and scattering of Ly $\alpha$ photons, respectively. Spin temperature is tightly coupled with gas temperature when $x_{\mathrm{tot}} \equiv x_{c}+x_{\alpha} \gtrsim 1$ and relaxes to $T_{\gamma}$ when $x_{\text {tot }} \ll 1$.

The spin temperature can be determined as follows:

\begin{equation}
T_{\mathrm{spin}}^{-1} \simeq \frac{T_{\mathrm{CMB}}^{-1}+\sum_{i} x_{i} T_{\mathrm{b}}^{-1}}{1+\sum_{i} x_{i}}.
\end{equation}

Brightness temperature (radiative transfer) could also be find by the following formula:

\begin{equation}
T_{\text {bright }}=T_{\text {spin }}\left(1-e^{-\tau}\right)+T_{\mathrm{CMB}} e^{-\tau}.
\end{equation}


Brightness temperature contrast is therefore determined as follows:
\begin{equation}
T_{21} \equiv \delta T_{\text {bright }}=\frac{T_{\mathrm{spin}}-T_{\mathrm{CMB}}}{1+z}\left(1-e^{-\tau}\right) \approx \frac{T_{\mathrm{spin}}-T_{\mathrm{CMB}}}{1+z} \tau.
\end{equation}


 
In figure \ref{The Cosmological 21cm Signal} the Cosmological 21 cm Signal is shown where one of the observable is
\begin{equation}
T_{21}(z) \approx \frac{T_{\mathrm{Spin}}-T_{\mathrm{CMB}}}{1+z} \sim 23 \mathrm{mK} \times x_{\mathrm{HI}}(z)\left[\left(\frac{0.15}{\Omega_{m}}\right)\left(\frac{1+z}{10}\right)\right]^{1 / 2}\left(\frac{\Omega_{b} h}{0.02}\right)\left(1-\frac{T_{\mathrm{CMB}}(z)}{T_{\mathrm{Spin}}(z)}\right).
\end{equation}


 

In the 21 cm line, two types of background radio sources play a critical role in probing astrophysics. CMB can be used as a radio background source, first and foremost. In this case, $T_{R}=T_{\mathrm{CMB}}$ and the $21 \mathrm{~cm}$ features can be viewed as spectral distortions of the CMB blackbody at appropriate radio frequencies (since the temperature fluctuations of the CMB are low $\delta T_{\mathrm{CMB}} \sim 10^{-5}$, the CMB appears uniformly bright). As with the CMB anisotropies, the distortion forms a diffuse background that can be studied across the whole sky. In order to build a $3 \mathrm{D}$ map of the observable Universe, observations at different frequencies probe different shells of the observable Universe. 

Radio loud point sources such as quasars are used as the background in the second situation. Sources in this situation will always emit more light than weak emission from diffuse hydrogen gas, $T_{R} \gg T_{S},$ so that gas is observed in absorption when compared to the source. Various distances away from the source produce lines that form a "forest" known as the "21 cm forest," like the Ly $\alpha$ forest. The high brightness of the background source allows the $21 \mathrm{~cm}$ forest to be studied with high frequency resolution so probing small scale structures $(\sim k p c)$ in the IGM. For useful statistics, a number of lines of sight are needed to reach different radio sources, prioritizing the discovery of high redshift radio sources.

We have a number of quantities that contain units of temperature, and many of them are not true thermodynamic temperatures. $T_{R}$ and $\delta T_{b}$ are radio intensity measurements. $T_{S}$ measures the relative occupation numbers of the two hyperfine levels. $T_{\alpha}$ is a color temperature describing the photon distribution in the vicinity of the Ly-$\alpha$ transition. There is no true thermodynamic temperature other than the CMB blackbody temperature  $T_{\mathrm{CMB}}$ and $T_{K}$.\\

\subsection{Collisional coupling}

As the gas density is high in the early Universe, collisions between particles can cause spin-flips in atoms of hydrogen and dominate the coupling there. Collisions between two hydrogen atoms, collisions between a hydrogen atom and an electron, and collisions between a hydrogen atom and a proton, are the three principal channels. Assuming '$i$' is a species, the collisional coupling is as follows:

\begin{equation}
x_{c}^{i} \equiv \frac{C_{10}}{A_{10}} \frac{T_{\star}}{T_{\gamma}}=\frac{n_{i} \kappa_{10}^{i}}{A_{10}} \frac{T_{\star}}{T_{\gamma}}.
\end{equation}

Collisional excitation rate is $C_{10}$ , $\kappa_{10}^{i}$ is the specific rate coefficient for spin deexcitation by collisions with species $i$ (in units of $\left.\mathrm{cm}^{3} \mathrm{~s}^{-1}\right)$.
The total coefficient of collision can be expressed as:

\begin{equation}
\begin{aligned}
x_{c} &=x_{c}^{H H}+x_{c}^{e H}+x_{c}^{p H} \\
&=\frac{T_{\star}}{A_{10} T_{\gamma}}\left[\kappa_{1-0}^{H H}\left(T_{k}\right) n_{H}+\kappa_{1-0}^{e H}\left(T_{k}\right) n_{e}+\kappa_{1-0}^{p H}\left(T_{k}\right) n_{p}\right],
\end{aligned}
\end{equation}

where $\kappa_{1-0}^{H H}$ is the scattering rate between hydrogen atoms, $\kappa_{1-0}^{e H}$ is the scattering rate between electrons, hydrogen atoms, and protons and hydrogen atoms scatter at a rate of $\kappa_{1-0}^{p H}$ .

Quantum mechanical calculations are needed to calculate collisional rates. On the basis of $T_{k}$ we have tabulated $\kappa_{1-0}^{H H}$ values, the scattering rate between electrons and hydrogen atoms $\kappa_{1-0}^{e H}$ was considered and the scattering rate between protons and hydrogen atoms $\kappa_{1-0}^{p H}$ was considered in \cite{Furlanetto}. This scattering rate can be fitted by the following fitting functions: A well-fitting HH scattering rate can be seen in the range $10 \mathrm{~K}<T_{K}<10^{3} \mathrm{~K}$ by 

\begin{equation}
\begin{aligned}
\kappa_{1-0}^{H H}\left(T_{K}\right) \approx 3.1 \times 10^{-11} T_{K}^{0.357} \exp \left(-32 / T_{K}\right) \mathrm{cm}^{3} \mathrm{~s}^{-1},
\end{aligned}
\end{equation}
and the scattering rate of e-H is well described by

\begin{equation}
\begin{aligned}
\log \left(\kappa_{1-0}^{e H} / \mathrm{cm}^{3} \mathrm{~s}^{-1}\right)=-9.607+0.5 \log T_{K} \times
\exp \left[-\left(\log T_{K}\right)^{4.5} / 1800\right] ,
\end{aligned}
\end{equation}

for $T \leq 10^{4} \mathrm{~K}$ and $\kappa_{1-0}^{e H}\left(T_{K}>10^{4} \mathrm{~K}\right)=\kappa_{1-0}^{e H}\left(10^{4} \mathrm{~K}\right)$.
The details of the coupling process become important during the cosmic dark ages, where collisional coupling dominates. The above calculations assume collisional cross-sections are independent of velocity; in fact, velocity dependence results in a non-thermal distribution for the hyperfine occupation. This effect can lead to a suppression of the 21 cm signal at the level of $5 \%$, although small, the signal measures 21 cm, which can be used for precision cosmology utilizing the 21 cm  signal from the dark ages. 
\\

\subsection{Wouthuysen-Field effect}

As a result of the resonant scattering of Ly-$\alpha$ photons, the photon spectrum around Ly-$\alpha$ frequency $\nu_{0}$ will have a local Boltzmann distribution with a color temperature equal to the kinetic temperature of hydrogen gas. This implies that the spin temperature $T_{s}$ of neutral hydrogen will directly be coupled to the kinetic temperature of hydrogen gas. Here is what's known as the Wouthuysen-Field (W-F) coupling. It is essential to estimate the redshifted 21 cm signal from the halos of the first generation of stars due to the substantial spin temperature deviation from the cosmic microwave background (CMB) temperatures ($T_{cmb}$).
It is not as simple as a simple description would suggest the physics behind the Wouthuysen-Field equation. The coupling could be written as \cite{Wouthuysen-Field Coupling}:

\begin{equation}
x_{\alpha}=\frac{4 P_{\alpha}}{27 A_{10}} \frac{T_{\star}}{T_{\gamma}},
\end{equation}

where $P_{\alpha}$ is the scattering rate of Ly-$\alpha$ photons. Here we have related the scattering rate between the two hyperfine levels to $P_{\alpha}$ using the relation $P_{01}=4 P_{\alpha} / 27,$ which results from the atomic physics of the hyperfine lines and assumes that the radiation field is constant across them. An atom of hydrogen scatters Ly-$\alpha$ photons according to the following equation:

\begin{equation}
P_{\alpha}=4 \pi \chi_{\alpha} \int \mathrm{d} \nu J_{\nu}(\nu) \phi_{\alpha}(\nu).
\end{equation}

The local absorption cross section can be calculated by using $\sigma_{\nu} \equiv \chi_{\alpha} \phi_{\alpha}(\nu)$, $\chi_{\alpha} \equiv\left(\pi e^{2} / m_{e} c\right) f_{\alpha}$ is the oscillation strength of the Ly $\alpha$ transition, $\phi_{\alpha}(\nu)$ is the Ly-$\alpha$ absorption profile, and in the background radiation field, $J_{\nu}(\nu)$ is the angle-averaged specific intensity, (by number$)$.

 
Wouthuysen-Field coupling, or the Wouthuysen-Field effect, occurs when the excitation temperature of neutral hydrogen is coupled to the Lyman-$\alpha$ (Ly-$\alpha$) radiation. When the Dark Ages ended and the epoch of reionization began, this coupling led to there being a difference in temperature between neutral hydrogen and the cosmic microwave background.
 
There are times when neutral hydrogen's thermodynamic equilibrium with the cosmic microwave background (CMB) prevails, but the hydrogen line cannot be observed because the CMB and neutral hydrogen are "coupled". The hydrogen line is only visible when the two temperatures differ, i.e. when they are decoupled.
 
Wouthuysen-Field coupling is a mechanism by which neutral hydrogen's spin temperature is coupled to Ly-$\alpha$ radiation, which decouples it from the CMB. In astrophysical sources such as stars and quasars, the Ly-$\alpha$ transition produces energy of approximately 10.2 eV, which is approximately 2 million times greater than the hydrogen line. As Ly-$\alpha$ photons are absorbed, Ly-$\alpha$ photons are re-emitted, and hydrogen can go into either one of two possible spin states. Electrons are redistributed between the hyperfine states in this process, allowing neutral hydrogen to be decoupled from CMB photons.

The coupling between the Ly-$\alpha$ photons and the hyperfine states depends not on the intensity of the radiation, but on the shape of the spectrum located in the vicinity of Ly-$\alpha$ transitions. It was first suggested by S. A. Wouthuysen in 1952 that this mechanism might affect the population of hyperfine states in neutral hydrogen, and later developed by B. George  Field in 1959.

Hyperfine levels are affected by the Ly-$\alpha$ photons depending on the intensity of the red and blue wings of the line, which reflects a very small difference in energy between the hyperfine states and the transition. A cosmological redshift of 6 will cause Wouthuysen-Field coupling to raise the spin temperature of neutral hydrogen above its CMB counterpart, which will produce hydrogen-line emission.\\

 \section{Current Global 21 cm Signal Experiments}
 Despite the newness of this method, there are only a few telescopes in use today. Another major project, the Square Kilometer Array, is in the planning stages. In April 2013, a research team using data collected by the Green Bank Telescope reported a detection that placed upper and lower limits on the 21 cm background \cite{global experiments}.
 
\subsection{Experiment to Detect the Global Epoch of Reionization Signature (EDGES)}
 
The Experiment to Detect the Global EoR Signature (EDGES) is a ground-based experiment that seeks to detect the 21 cm radiation from neutral hydrogen in the intergalactic medium (IGM) at redshifts $27 > z > 6$. The redshifts are caused by the formation of the first stars, galaxies, and black holes, as well as the reionization process of the IGM. The redshifts of the 21 cm radiation can be measured with radio frequency bands between 50 and 200 MHz. This mission serves as a pathfinder for the Moon-orbiting DARE mission and the NESS-funded Moon-based low-frequency cosmology telescope.

In 2007, EDGES began conducting measurements and has been placed at the Murchison Radioastronomy Observatory in Australia's desert to take full advantage of the low levels of artificial interference (RFI). The largest challenge in this measurement is to distinguish between the very weak 21 cm signal ($<500 mK$) and the strongly diffuse foregrounds between hundreds and thousands of Kelvin. Because of this, it is necessary to calibrate an instrument to a level of about 1 part in 10,000 in order to achieve sufficient sensitivity for detection. EDGES has enabled the tightest constraints to date on the global 21 cm signal due to its sensitivity being comparable to this requirement
\cite{colorado}.

 
 
\subsection{The Large-Aperture Experiment to Detect the Dark Ages (LEDA)}

Large-Aperture Experiment to Detect the Dark Age (LEDA) will detect the predicted O(100) mK sky-averaged absorption of the cosmic microwave background by hydrogen in the neutral pre- and intergalactic medium just after the cosmological Dark Age. When 'Cosmic Dawn' occurs, the spectral signature will be associated with emergence of a diffuse Ly-$\alpha$ background from starlight. Recently, \cite{Bowman et al} reported detecting an amplitude of 530 mK, centered at 78 MHz, of the predicted absorption in this signal. It is imperative that LEDA independently validates this finding. The LEDA system is part of the Owens Valley Radio Observatory Long Wavelength Array, which consists of a station correlator, five redundant dual polarization radiometers well positioned apart, and back-end electronics. In an attempt to calibrate, the radiometers operate in a 30-85 MHz frequency band $(16<z<34)$ as part of a larger interferometric array  \cite{ui.adsabs.harvard.edu, tauceti}.

 
\subsection{The Hydrogen Epoch of Reionization Array (HERA)}

Hydrogen Epoch of Reionization Array (HERA) attempts to measure the emissions of the primordial intergalactic medium (IGM) throughout cosmic reionization (z = 6 - 12). In addition, it is also possible to study earlier eras of our Cosmic Dawn ($z\sim 30$). As stars and black holes heated and ionized the IGM during these periods, 21 cm emission fluctuated. It will use the properties of the first galaxies, the evolution of large-scale structure, and the early sources of heat to constrain the timing and morphology of reionization. The HERA instrument will consist of 350 elements in South Africa and will be a 14-m parabolic dish interferometer that observes with a frequency range of 50 to 250 MHz ,\cite{astro.phy.cam.ac.uk, reionization}. 
 
\subsection{The Precision Array to Probe the Epoch of Reionization(PAPER)}

The Precision Array for Probing the Epoch of Reionization (PAPER) is a low-frequency radio interferometer used to measure the occurrences of the first stars and galaxies around 500 million years after the Big Bang. Researchers measured the power spectrum of intergalactic medium fluctuations introduced by the first luminous sources at high redshifts (z = 7 - 12) by mapping the intensity of hydrogen emission at 21 cm.

With foregrounds five orders of magnitude brighter than the background standards for detecting reionization, the PAPER project takes an incremental engineering approach, optimizing each component in the array in a staged process to minimize potential problems with subsequent calibration and analysis. With the help of this staged approach, we can systematically address observations challenges resulting from very wide field, high dynamic range imaging within large bandwidths under the influence of transient terrestrial interference
\cite{berkeley}. 


\subsection{Probing Radio Intensity at high-Z from Marion(PRIZM)}

Using globally averaged redshifted 21 cm emission, a new experiment examines the cosmic dawn absorption feature. PRIZM is among a number of previous and current global signal experiments.
This experiment uses two dual-polarization radiometers with centers frequencies of 70 and 100 MHz; their combined frequency range is 30 - 200 MHz. An overview of the PRIZM signal chain for a single polarization can be seen in Figure 15. As with the original SCI-HI experiments, PRIZM radiometer cells use an antenna design developed for the HIBiscus experiment 
\cite{Probing Radio Intensity}. 


\subsection{Measurement of Background Radio Spectrum (SARAS)}

All-sky 21 cm signals are distorted by 10's to about 100 mK, and this is present in the cosmic radio background at 30-200 MHz as a trace additive component. The galactic and foreground brightnesses of these long wavelengths of contaminating radio sky are between 100 and 10,000 K. Furthermore, systematics from within radio-frequency spectrometers and outside interference (RFI) is a major challenge in the task of designing the detection system and developing the algorithms required. Towards this key science goal we are working on the experiment SARAS (Shaped Antenna measurement of the background RAdio Spectrum), as part of this program, there is a continuous development, deployment, testing, and improving of a spectral radiometer that is based on an independent frequency antenna, self-calibrating receivers, and a broadband precision digital spectrometer.

As a correlation spectrometer, SARAS was designed to make measurements of the cosmic radio background and faint features in the sky spectrum at long wavelengths caused by gas redshifted to 21 cm at the time of reionization. System parameters are calibrated by measuring the difference in antenna temperature and an internal reference and a complex switching scheme to cancel systematics. In ref. \cite{Patra et al 2013}, details of the architecture and data analysis procedure used in SARAS are given.

In the observed spectrum, the foregrounds and systematics are dominant, and the antenna and receiver characteristics link the foregrounds and systematics to the detector with complex transfer functions that are also different for different components of the response. The reionization signal will be confused by all of this. To minimize instrument signatures, it is important to pay attention to the system design, characterization, and calibration. Therefore, we are concentrating on developing algorithms and methods for properly calibrating the system, minimizing systematics, and modeling the factors contributing to the system response with mK accuracy to discover the cosmologically redshifted 21 cm signal.

In addition to continuous improvements in antenna design, analog and digital receiver design, we test potential observing sites to choose ones with relatively little radio frequency interference (RFI) \cite{rri, SARAS}. 


\subsection{Broadband Instrument for Global Hydrogen Reionisation Signal (BIGHORNS)}

The Broadband Instrument for Global Hydrogen ReioNisation Signal (BIGHORNS) is a total power radiometer designed to detect reionisation signatures in low-frequency radio waves (70 - 300 MHz).

An Epoch of Reionization (EoR) represents a period of time when the previously neutral intergalactic medium became ionized by the first luminous sources. This transition line is extremely difficult to observe for this period, but its detection has been considered a vital objective for radio astronomy. As a result, a variety of detection systems are being investigated, from single antennas such as BIGHORNS to large interferometer arrays such as the MWA.

A conical log spiral antenna, known as BIGHORNS, was deployed at the Murchison Radio-astronomy Observatory (MRO) in October 2014, and has been collecting data ever since. In ref. \cite{BIGHORNS}, present details of the original biconical antenna system and its upgrades has been published.

BIGHORNS was also instrumental in assessing the level of radio frequency interference (RFI) at the future site for the low-frequency component of the Square Kilometre Array \cite{BIGHORNS}. 


\subsection{Sonda Cosmol ogica de las Islas para la Detecci on de Hidr ogeno Neutro (SCI-HI)}

The 'Sonda Cosmologica de las Islas para la Deteccion de Hidrogeno Neutro' (SCI-HI) experiment  is a 21 cm brightness temperature spectrum experiment that studies the cosmic dawn ($z\sim15 - 35$). It is a collaboration between Carnegie Mellon University (CMU) and Instituto Nacional de Astrofísica, Óptica Electrónica (INAOE) in Mexico. This experiment was launched in June 2013 on the small island of Guadalupe, which is about 250 km off the coast of Baja California in Mexico. The first observations from this deployment have placed the first constraint on the 21 cm all sky spectrum around 70 MHz ($z\sim20$), \cite{Voytek et al (2014)}. There are a great deal of neutral hydrogen (HI) molecules found in the intergalactic medium (IGM). In this case, the HI can be seen through the 21 cm (1.4 GHz) spectral line as a result of hyperfine structure. A wavelength of this spectral line extends at a rate defined by the redshift $z$ with the expansion of the universe, producing a signal traceable through time.

It turns out that the strength of the 21 cm signal in the IGM is only affected by a few variables: the temperature and density of the IGM, the amount of HI in the IGM, the UV energy density in the IGM, and the redshift. 21 cm measurements, therefore, provide us with insights into the history and structure of the IGM. During the cosmic dawn before reionization, the SCI-HI experiment analyzes the spatially averaged 21 cm spectrum to observe the evolution of the IGM over time. The SCI-HI experiment aimed to constrain the initial data. However, the preliminary data centered around a narrow band between 60 and 85 MHz. Radio frequency interference (RFI) and instrumental difficulties limited radio transmission in the FM radio band ($\sim 88 - 108$ MHz). A new deployment of SCI-HI is currently in progress and we plan to have another soon. This deployment will take place at Socorro and Clarion, two islands farther from mainland Mexico than Guadalupe
\cite{ui.adsabs.harvard.edu2}.


\subsection{ Square Kilometre Array (SKA)}

With a square kilometre of collecting area, the Square Kilometre Array (SKA) will be the world's most sensitive radio telescope. A multi-nation project, the SKA design and construction involves 15 countries. It was decided in 2012 to select two sites in Australia and one in South Africa, as well as to have headquarters in the United Kingdom.

A technique known as aperture synthesis will be used to combine the signals received from thousands of small antennas spread over thousands of kilometres to simulate an extremely sensitive and angular-resolution radio telescope. There will also be sub-arrays of the SKA that will have very large field-of-view (FOV), allowing for surveying very large areas of the sky simultaneously. Using the SKA, researchers will be able to analyze a wide range of questions in astrophysics, fundamental physics, cosmology, and particle astrophysics, in addition to expanding the observable universe. A key purpose of the SKA is to provide observations from the so-called Dark Ages (between 300,000 and 600,000 years after the Big Bang when the universe first became sufficiently cool for hydrogen to become neutral and decoupled from radiation)  \cite{https://www.skatelescope.org/layout/}.

Jodrell Bank Observatory in Cheshire, England, hosted the SKA's headquarters, while the telescopes will be installed in Australia and South Africa  \footnote{https://www.industry.gov.au/policies-and-initiatives/co-hosting-the-square-kilometre-array}. It is important that SKA telescope sites are unpopulated areas with very low levels of radio interference caused by humans. South Africa, Australia, Argentina, and China were among the first four proposed sites. 
\footnote{http://www.ncra.tifr.res.in:8081/SKA-India/}.



 
\subsubsection{Cosmology with other hyperfine structure lines and SKA}

Hyperfine structure lines can also be found in other atomic species that may be useful for cosmological studies with the SKA. $\lambda=91.6 \mathrm{~cm}$ represents the hyper-fine line of neutral deuterium. This line is more challenging to detect than hydrogen's $21 \mathrm{~cm}$ line due to its longer wavelength and also to higher abundance $[\mathrm{D} / \mathrm{H}]$ relative to hydrogen; however, it provides the most accurate measurement of primordial abundance since it is free of contamination by processes of structure formation at lower $z$. Using this method, we can indirectly determine the baryon-to-photon ratio $\eta=n_{\mathrm{b}} / s .$ Big bang nucleosynthesis $(\mathrm{BBN})$ provides the only known source of deuterium, which in turn lets us determine the cosmic baryon abundance, $\Omega_{\mathrm{b}} h^{2}$ \cite{skatelescope}.\\

\subsection{Murchison Widefield Array (MWA)}

Among the few experiments designed to detect the statistical signal from the Epoch of Reionisation, the MWA EoR is one of the more prominent ones. In each of these experiments, the challenges, such as removing from the foreground, have become more understood. Through the MWA Collaboration's expertise and understanding of the telescope array, the end-to-end pipelines, the ionospheric conditions, and the foreground emissions, the collaboration has developed expertise and understanding. It has been possible to detect the theoretically predicted EoR signal based upon the collected data. There are periodic limit updates, but the detection is still several orders of magnitude away. Throughout this paper, recent progress is summarized, along with future directions.

The next decade will hold unprecedented opportunities for understanding the Cosmic Dawn, the period when the first galaxies and stars were formed. A radio interferometer that has been using the MWA to detect redshifted 21 cm line emission from neutral hydrogen gas in the intergalactic medium (IGM) during this period may be one of the first to detect these emissions. In order to optimize the array's capability to detect brightness temperature variations in the 21 cm emission during the EoR, it was designed to have the redshift range $6 < z < 10$ optimized. It is during the EoR that the first luminous sources begin to ionize the primordial neutral hydrogen. Through measurements of the power spectrum and other statistical properties of the fluctuations, the MWA is capable of detecting the presence of large ionized bubbles that form during reionization to a significance level of 14 sigma \cite{Beardsley et al}. It will be used as a test bed to demonstrate and develop techniques to subtract the bright radio foregrounds that obscure the 21 cm background in order to implement this objective \cite{MWA, mwatelescope}.\\ 

\subsection{Low-Frequency Array (LOFAR)}

LOFAR, the Low-Frequency Array, is a new-generation interferometer being built in the Netherlands and throughout Europe. Through the use of a phased-array, LOFAR provides access to the low-frequency range from 10–240 MHz, which has been largely unexplored. LOFAR stations are being constructed throughout the Netherlands, spread out from the village of Exloo in the northeast. There are also five other stations deployed in Germany, and three stations have been built in France, Sweden, and the United Kingdom. By using digital beam-forming techniques, LOFAR can be made more agile and able to carry out many simultaneous observations at once. LOFAR achieves amazing angular resolution and sensitivity in the low-frequency radio regime thanks to its dense core array and long interferometric baselines. As an observatory operated by the International LOFAR Telescope (ILT) foundation, LOFAR facilities are available to astronomers throughout the world. It is, in fact, one of the first radio observatories to provide its users with fully calibrated science products via automated processing pipelines.

Due to LOFAR's new capabilities, techniques, and operational modes it is an important stepping stone to the SKA. In the very distant Universe $(6<z<10)$, LOFAR can identify the results of hydrogen reionization. As the first stars and galaxies form at the epoch of the so-called "dark ages", this phase change will occur at the same time. When reionization begins, the 21 cm line of neutral hydrogen at $1420.40575 \mathrm {MHz}$ will enter the LOFAR observing window. There is a factor of $1 /(z+1)$ difference in frequency observed today \cite{LOFAR: The LOw-Frequency ARray}.\\ 



\section{Using 21 cm line radiation to Map the Galaxy}

The 21 cm emission line intensity varies with the density of atomic hydrogen centered along a line of sight. Radio emission Doppler-shifted can be converted into distances to hydrogen clouds by using the rotation curve.

Rotation curves are plots of the orbital velocity of clouds around the galactic center against their distance from the galaxy center. When used in this context, 'rotation' refers to a galactic disk's motion as a whole. It appears as though a disk of stars and gas clouds is spinning. The gas clouds are presumed to revolve on nearly circular orbits in the disk's plane. In 1927, Jan Oort discovered that stars closer to the galactic center complete more of their orbits in a given time than those farther out. Differential rotation describes the difference between the angular speeds of different parts of the galactic disk.

By observing the Doppler velocities of hydrogen gas along different lines of sight, the rotation curve can be determined. Hydrogen at varying distances from the galactic center will contribute to the 21 cm emission, which will have a different Doppler shift with respect to us. Gas clouds inside the orbit of the sun will emit some of the radiation because they move at slightly faster angular speeds than the sun. Small redshifts are expected. Gas moving at the greatest angular speed will contribute the most to the total emission coming from the closest region to the galactic center,  \cite{http://www.astronomynotes.com/ismnotes/s3.htm},\cite{21 cm Intensity Mapping}.\\
 

\subsection{Three dimensional ‘Intensity Mapping’ of the Universe}

Up until now, the primary method for determining the large scale structure of the universe was through galaxy redshift surveys, that is by isolating millions of individual galaxies, recording their spectra, and measuring their redshifts. The primary technique used for this is optical spectroscopy. In total, optical surveys have mapped just under 1\% of the observable universe.\\

\subsection {Measuring Baryon Acoustic Oscillations with the 21 cm line of hydrogen}

In order to analyze the expansion history, the Baryon Acoustic Oscillations method (BAO) is proposed. The first 380,000 years of expansion were characterized by extremely ionized cosmic matter and, as a consequence, acoustic waves could propagate across the universe. Wave crests from the ionized era left behind density enhancements, but acoustic waves no longer propagated at its end
\cite{BAO}.

According to simulations, the 21 cm intensity map technique allows very precise constraints to be applied to the equation of state for dark energy. The survey was sensitive due to its large volume. \\
\subsection{Foreground Emission}

Both galactic and extragalactic foregrounds are strongly dominated by synchrotron emission at these wavelengths. A nearly power law distribution of electron cosmic ray energies is what is expected to produce a nearly power law distribution of synchrotron radiation foreground. According to recent estimates, the foreground spectrum of a synchrotron must always be smooth, regardless of how the electron energies are distributed. Under the extreme assumption that electrons are monoenergetic, it is still more smooth by more than ten orders of magnitude than is required for successful subtraction. Because the synchrotron emission process does not require spectral features, the 3-D 21 cm intensity signal stands out due to its lack of spectral features \cite{intensity mapping}.

\section{21-cm Emission Around The Sun}

\subsection{Brightness Temperature Of The Sun}

Using the interstellar hydrogen emission, one can measure the sun's brightness temperature at 21 cm. The substate of high energy hydrogen is found in two-thirds of hydrogen atoms at 100 K, the temperature of the interstellar medium. Despite the fact that only one hydrogen atom emits 1420.4 MHz radiation every $10^{7}$ years, enough hydrogen atoms transition for measurable quantities of this radiation to occur. It results from the vast numbers of hydrogen atoms that are found within any galactic line of sight.

When the antenna temperature is calculated from the solar temperature, it is the solar temperature plus the area surrounding the sun. The sun radiates from a surface $\pi\left(L\psi_{\operatorname{sm}}^{\mu}\right)^{2}$ and the view area is $x\left(L\psi_{\text {view}}\right)^{2}$.
The actual brightness temperature of the sun is

\begin{equation}
T_{b}\left(\nu_{0}\right)=\frac{T_{a n t}}{\eta}\left(\frac{\psi_{v i e w}}{\psi_{s u n}}\right)^{2} ,
\end{equation}

where $\psi_{s u n}=32^{\prime}=0.533^{\circ} \pm 0.018^{\circ}$ is the angular diameter of the sun seen from the earth and $\psi_{\text {view}}=$ $6.41^{\circ} \pm 0.03^{\circ}$ is the viewing angle of the telescope. With these equations, the brightness temperature of the sun can be determined $T_{b}=(4.5 \pm 0.3) 10^{4} K$, \cite{http://joans.io/papers/21cm.pdf}.\\

\subsection{The Radio Brightness Distribution On The Sun At 21 Cm From Combined Eclipse And Pencil-Beam Observations}

Observations of the emission of radiation from the sun at decimetre wavelengths at that time showed that it fluctuated with a period comparable to the rotation period of the Sun, but was normally steady over short periods. These observations were made with aerials whose beamwidths to half-power points were much greater than the angular size of the Sun. The low resolution devices that we use to observe the Sun will henceforth be referred to as "radiometers".

At a wavelength of $10.7 \mathrm{cm}$ wavelengths, Covington in 1947 observed a partial eclipse of the Sun. A sharp decrease in flux density was found simultaneous with the occultation of a large visible sunspot group, suggesting the presence of a localized radio source. According to a statistical analysis of the day-to-day variation in radio emission from the sun at various wavelengths, this is true, (\cite{Pawsey and Yabsley 1949} and \cite{ Denisse 1949}). In the research \cite{Krishnan Labrum}, it was found that the radio flux at a specific wavelength was always strongly correlated with the projected sunspot area, which is the main component of solar radiation. By fitting a regression line to the data, we could distinguish between the two components. In addition to the extrapolated flux value for an unspotted disk, the variable solar radiation component was called the "slowly varying component" since its ratio was proportional to sunspot area.

The quiet-Sun component could be explained by thermal radiation from the entire Sun. There were however fewer evidences that the slowly varying component and sunspot area were associated. Due to the fact that radio emission at decimetre wavelengths originates at levels thousands of kilometres above the photosphere, the associations between the slowly varying component and sunspot area were less clear. As in ref. \cite{Pawsey and Yabsley 1949} confirmed this with eclipse observations showing in some cases radio sources were present where old sunspot groups were no longer visible. There seemed to be a need to search for long-lasting solar features that could emit radio waves due to their association with sunspot groups. According to \cite{Waldmeier and Muller 1950}, coronal condensations may be responsible for the slowly varying component.

As in ref. \cite{Christiansen Warburton Davies} been discovered that radio sources and astrophysical plages were always associated, and when the radio sources were resolved, the astrophysical plages matched their sizes. The plage areas and sizes of radio sources show a strong correlation, as confirmed by \cite{Christiansen and Mathewson (1958)}.

Similarly, for 21 cm wavelengths, radio bright regions or "radio plages" have a shape and size that are comparable to those of related chromospheric plages. They have given an illustration that shows the correspondence in shape very well. 

The quiet-Sun component has been studied at sunspot minimum and maximum in high resolution. According to references \cite{Christiansen and Warburton (1955)} and \cite{Christiansen and Warburton (1955)2}, the distribution of brightness on the quiet Sun at a wavelength of 21cm displayed quadrant rather than circular symmetry; the limb of the Sun was bright at the equator but not the poles. The limbs on the equator reached a temperature of $6 \cdot 8 \times 10^{5} \mathrm{K}$ , while the temperature at the centre of the Sun's disk was $4 \cdot 6 \times 10^{5}$ $\mathrm{K}$  \cite{http://adsabs.harvard.edu/full/1961AuJPh..14..403K}.\\

    {sunplot.pdf}


\section{Probing for dark matter and dark energy with 21 cm signal}
To know the 21 cm observation application  in cosmology, first we should know how dark side of universe appear in cosmological observation.

\subsection{Milky Way Galaxy Rotation Curve And Dark Matter}
If all of the mass in the galaxy is visible, and if the galaxy is radially symmetric, the Keplerian curve can be calculated. Considering the currently accepted model of galaxy formation is a dense bulge surrounded by spiral arms, the latter is true at large radii. A large radius brings out the effect of a uniform disk. To calculate the rotation curve, Newton's law and universal gravitation can be applied as follows:

\begin{equation}
\frac{m v^{2}}{r}=\frac{G m M}{r^{2}} \Longrightarrow v(r)=\sqrt{\frac{G M}{r}}.
\end{equation}

The universal gravitational constant G is equal to $10^{42}$kg, and the visible mass M equals the visible mass of the galaxy.
Based on the plot of the rotation curve, it is evident that the experimental data are contrary to Keplerian prediction. Accordingly, since the Keplerian curve is predicated upon a visible mass of the galaxy, this implies that there is less visible mass in the galaxy. According to the current hypothesis, dark matter explains the difference between the two curves by being invisible ,\cite{http://joans.io/papers/21cm.pdf}. 

    {RotationCurveoftheMilkyWayGalaxy.pdf}

\subsection{A brief review on dark matter}

Approximately 85\% of the universe's matter consists of dark matter and it accounts for about a quarter of its total mass-energy density, or $2.241 \times 10^{-27} \mathrm{~kg} / \mathrm{~m}^{3}$. Many astrophysical observations indicate the existence of dark matter, including gravitational effects that cannot be explained by other theories of gravity if there is more matter present than can be seen. Most scientists believe that dark matter has a strong influence on the structure and evolution of the universe since it is abundant in it. Due to the fact that dark matter is invisible, it cannot be detected by electromagnetic fields, meaning that it cannot absorb, reflect, or emit electromagnetic radiation.

According to calculations, if galaxies did not contain dark matter, they would fly apart, wouldn't have emerged or wouldn't move as they do. As well as gravitational lensing and the CMB, astronomers have detected observable galaxies and observed galaxy formation and evolution, as well as the motion of galaxies within galaxy clusters. In $\Lambda$CDM cosmology dictates that the total mass and energy of the universe consists of 5\% ordinary matter and 27\% dark matter, and 68\% of a mysterious form of energy called dark energy. 

Unless dark matter interacts with ordinary baryonic matter, gravity must be its major interaction with the matter. Dark matter hasn't been observed directly, but if it does exist, it has a minimal effect on baryonic matter. Unlike baryonic matter, dark matter could consist of particles that are yet to be discovered. It's likely dark matter is made up of a new type of elementary particle that has yet to be discovered, for instance weakly interacting massive particles (WIMPs). Although many experimental efforts have been made to detect and study dark matter particles directly, none have been successful. According to its velocity (more specifically, its free streaming length), dark matter can be classified as "cold", "warm", or "hot". A cold dark matter scenario is currently preferred, in which structures occur as particles accumulate gradually.

Scientists generally accept the existence of dark matter, but certain observations that do not fit the standard theories, such as those that refute modified Newtonian dynamics and tensor-vector-scalar gravity, have led some to contemplate modifying the standard laws of general relativity \cite{modifiedgravity}. Using these models, it is attempted to explain all observations without adding non-baryonic matter.\\

\subsection{21 cm signal and Dark Matter}

All observations can be explained with the aid of these models, without including any non-baryonic matter. The corresponding energy input resulting from the decay or annihilation of dark matter (DM) affects the hydrogen kinetic temperature and ionized fraction and impacts the Ly-$\alpha$ background. The 21 cm signal can then be used to constrain the proposed DM candidates, and we choose three of the top three:

(i) 25-keV decaying sterile neutrinos, 
(ii) 10-MeV decaying light dark matter (LDM) and 
(iii) 10-MeV annihilating LDM. 
Even though the DM effects are much smaller than previously documented (because we model the energy transfer from the DM to the gas more physically), we conclude that the combination of observations of the 21 cm background and its gradient should allow constraints at least on LDM candidates \cite{constraining dark matter, indico}. 




It is believed that primordial density perturbations smaller than the free-streaming length-scale get dampened in the early Universe. At these scales, dark matter particles stream out of overdense regions into underdense regions, while fluctuations on scales larger than this one remain unchanged. An individual dark matter particle can be defined as having traversed a free streaming scale before the density perturbations become significant. Based on the calculation by \cite{kolb}:

\begin{equation}
\lambda_{\mathrm{fs}}=\int_{0}^{t_{\mathrm{mre}}} \frac{v(t) \mathrm{d} t}{a(t)} \approx \int_{0}^{t_{\mathrm{nr}}} \frac{c \mathrm{~d} t}{a(t)}+\int_{t_{\mathrm{nr}}}^{l_{\mathrm{mre}}} \frac{v(t) \mathrm{d} t}{a(t)},
\end{equation}
where $t_{\text {mre }}$ indicates the epoch of matter-radiation equality, and $t_{\mathrm{nr}}$ is the onset of non-relativistic behaviour of dark matter particles. As well, we have benefited from the fact that in relativistic domain, $v(t) \sim c .$ In the non-relativistic regime as $v(t) \sim a(t)^{-1}$ and during the radiation-dominated era, $a(t) \propto t^{1 / 2}$. In other words, the first equation ends up being:

\begin{equation}
\lambda_{\mathrm{Is}} \sim \frac{2 c t_{\mathrm{nr}}}{a_{\mathrm{nr}}}\left[1+\log \left(\frac{a_{\mathrm{mre}}}{a_{\mathrm{nr}}}\right)\right].
\end{equation}

Thus, increasing the time at which the nature of dark matter particles was relativistic, i.e. $t_{\mathrm{nr}},$. The free streaming scale $\lambda_{\mathrm{fs}}$ can be increased. Due to free streaming, we can calculate that a halo whose formation is suppressed has a mass of:

\begin{equation}
M_{\mathrm{fs}}=\frac{4}{3} \pi\left(\frac{\lambda_{\mathrm{Is}}}{2}\right)^{3} \bar{\rho}.
\end{equation}

As a result of CDM, dark matter particles become non-relativistic as they decouple $t_{\mathrm{nr}} \sim t_{\mathrm{dec}},$ meaning the free streaming length becomes negligible. In other words, it does not delete the initial perturbations on small scales. In CDM cosmology, therefore, galaxies form first, and galaxy clusters occur later. It has been confirmed in several simulations, \cite{Diemand Moore 2011} and \cite{Frenk White 2012}. A model like this is, however, inconsistent with small scale observations.

The WDM model's particles become non-relativistic later than the CDM model, and the $t_{\mathrm{nr}}$ is somewhere in between the two scenarios $t_{\mathrm{dec}}<t_{\mathrm{nr}}<t_{\mathrm{mre}} .$ The WDM model is therefore able to suppress density perturbations at small scales because the free streaming scales for dark matter particles are larger than for CDM particles. A bottom-up structure is created at scales above  $\lambda_{\mathrm{fs}}$ and structure formation occurs at scales much less than $\lambda_{\mathrm{fs}}$ via a top-down approach  \cite{Signs of Dark Matter at 21-cm}. The result is also supported by various WDM simulations \cite{Bode et al 2001}, \cite{Schneider et al 2012} and \cite{Viel et al 2013}.

According to EDGES's recent measurements of the global 21 cm spectrum, the strong absorption signal starts about $z = 17$. The excess over the expectation is estimated to be $3.8 \sigma$.

The idea behind this new measurement is intriguing. As the first stars are born around $z \simeq 20$, the cosmic gas in the universe is at its coolest before it is heated by X-ray radiation. The cosmic dawn occurs in this epoch (roughly $15 \leqslant z \lesssim$ 35 ). As it is usually interpreted, the hydrogen 21 cm transition measurement provides a unique indication of hydrogen temperature. The gas's temperature drops below that of the radiation as it decouples from the CMB at around $z \simeq 200$. There may be measurable effects of modifications to the cosmic history during that period on the 21 cm absorption spectrum corresponding to that period.

Dark matter  is at its coldest phase at the same time, having not yet been sparked up by non-linear gravitational collapse. Baryon elastic scatterings at approximately or prior to that time could cool the gas, which might affect the 21 cm absorption spectrum. This concept was further considered in and analysed in showing that light (sub-GeV) dark matter can leave the desired imprint once Ly-$\alpha$ radiation turns on, providing an exciting explanation of the phenomenon. These interactions must, however, compete with Compton scatterings, which serve as a coupling device between the gas and the CMB radiation, and as a consequence must be quite strong. The strongest possible evasion of present-day, astrophysical and cosmological constraints. In particular, the BBN and CMB bounds on the number of relativistic degrees of freedom, the CMB anisotropy 5-th force experiments, and the stellar cooling bound on DM interactions and millicharge. In order for these interactions to occur, a light degree of freedom will be required, allowing Rutherford-like scattering with baryons. In the two-particle interaction, $\sigma=\hat{\sigma} v_{\mathrm{rel}}^{-4}$, where $v_{\text {rel }}$ is the relative velocity.

As in \cite{Tashiro},\cite{Mu_oz_2015},\cite{Barkana}, approach used for DM-hydrogen interaction, as well as CMB studies, is a model-independent one, where the velocity dependence in the cross section is assumed without stating its origin. Recent advancements are compelling enough to ask whether there exists a particle physics model which can account for both the EDGES observations and be consistent with current limits. To address this issue, we argue that the dominant component in DM cannot explain the cooling of hydrogen by the dominant component in DM.

The de Broglie wavelength of a DM particle with $m_{\mathrm{DM}} \lesssim \mathrm{GeV}$ may help us to understand this, and always greater than the atomic Bohr radius, $a_{0}=\left(\alpha m_{e}\right)^{-1} . \quad$ Therefore $\mathrm{DM}$ interacts with the hydrogen atom as a whole. In order to have a $v_{\text {rel }}^{-4}$ enhanced scattering cross-section, its overall charge must not vanish. However, that very same property implies that this mediator induces a long range force. Indeed, the mediator mass, $m_{\phi},$ must be smaller than the typical momentum transfer in order to induce a $1 / v_{\mathrm{rel}}^{4}$ enhancement. Since the relative velocity at the cosmic dawn is $v_{\text {rel }} \lesssim 10^{-6}$ one finds $m_{\phi} \lesssim \mathrm{keV}$ for $m_{\mathrm{DM}}<\mathrm{GeV}$.

In the case of a light mediator of this type, the constraints from 5th-force experiments are extremely strong (except for the upper allowed region, where stellar cooling constraints are stronger), eliminating the possibility of cooling via neutral hydrogen interaction. There are similar constraints for any mediator where heavier atoms are charged.

Due to the small residual fraction of free electrons and protons, there can still be Rutherford-like interactions between the DM and the gas. In comparison with DM-hydrogen, the interaction rate required is about three orders of magnitude larger. Early in the universe, proton interactions dominate the interaction rate, which implies a larger DM electron cross section. The DM-gas scattering cross sections will be probed with upcoming direct detection experiments With this discussion, two options remain for the mediation of the scattering: A $U(1)_{D}$ gauge boson (hidden photon) that is kinematically mixed with the Standard Model (SM) photon, or with the SM photon itself with a DM millicharge. Massless limits imply millicharged DM under electromagnetism, however the two theories are not equivalent: hidden photons can cause DM self-interactions, whereas DM in millicharge is strongly constrained by the galactic disk. \\

\subsection{Dark Cooling}

It is possible for dark matter to cool the spin temperature in two ways:
first, the hotter baryonic gas can be cooled by the darker matter fluid by scattering off gas particles. After the spin temperature couples with the gas temperature, the gas temperature decreases.
Second, DM can drain spin temperatures directly by, for example, directly spin-flip interactions (for sufficiently light DM) or through induced bosonically enhanced emissions.

It was in \cite{Tashiro} that dark matter cooling was first realized,
and further studied in 
\cite{Mu_oz_2015}
and
\cite{Barkana}.

The physics behind the phenomenon is summarized below. The interaction between DM and the gas should cool it down since DM is much colder. Under certain conditions, the predicted relative bulk velocity between the two gases dissipates and causes the gas to heat up. The Boltzmann equations can be used to illustrate the competing effect of temperature and velocity evolution:

\begin{equation}
\frac{d T_{\chi}}{d \log a}=-2 T_{\chi}+\frac{2}{3} \frac{\dot{Q}_{\chi}}{H},
\end{equation}

\begin{equation}
\frac{d T_{\mathrm{gas}}}{d \log a}=-2 T_{\mathrm{gas}}+\frac{\Gamma_{C}}{H}\left(T_{\mathrm{CMB}}-T_{\mathrm{gas}}\right)+\frac{2}{3} \frac{\dot{Q}_{\mathrm{gas}}}{H},
\end{equation}

\begin{equation}
\frac{d v_{\mathrm{rel}}}{d \log a}=-v_{\mathrm{rel}}-\frac{D\left(v_{\mathrm{rel}}\right)}{H}.
\end{equation}

Here $a$ indicates the scale factor, $H(z) \simeq \sqrt{\Omega_{m}} H_{0}(1+z)^{3 / 2}$ is the Hubble parameter during matter domination, $v_{\text {rel }}$ represents the relative velocity between DM and the gas, $\Gamma_{C}$ is the Compton scattering rate, and $D\left(v_{\text {rel }}\right)$ is the drag term that accounts for the relative velocity change due to DM gas interactions. From the scattering of the DM with the gas, $\dot{Q}_{\chi, \text { gas }}$ is the heat transfer per unit time.

The original calculations only took into account interactions with hydrogen. Free electrons and protons, however, are incredibly important, as we discussed in the introduction and will discuss below, due to the severe constraints on mediators which can cause velocity-enhanced DM hydrogen interactions. The terms describing interactions with hydrogen, helium, electrons and protons are carefully analyzed. The different contributions to heat transfer are thus described as follows

\begin{equation}
\dot{Q}_{\text {gas }}=\sum_{I=\{\mathrm{H}, \mathrm{He}, \mathrm{e}, \mathrm{p}\}} \dot{Q}_{\mathrm{gas}}^{I} .
\end{equation}

An approximate description of each contribution can be given as follows: $\dot{Q}_{\text {gas }}^{I} \sim x_{I} \Gamma^{I} \Delta E_{I}$ with $\Gamma^{I} \sim n_{\chi} \sigma^{I} v_{\mathrm{rel}}$ and $\Delta E^{I} \sim \mu_{I} v_{r e l}^{2},$ ( $\mu_{I}$ represents the reduced mass of the corresponding DM-gas component, $x^{I} \equiv n_{I} / n_{\mathrm{H}}$ and the DM number density is $n_{\chi}$). That leaves us with just one rationale for the need for large cross sections: To efficiently cool down the gas, the rate of DM-gas interactions must be comparable to the heating caused by Compton scattering. $\Gamma_{C} .$ As such, by requiring $\dot{Q}_{\text {gas }} \sim \Gamma_{C} T_{\text {CMB }}$ it is possible to estimate the required cross section $\hat{\sigma}^{I}$ that is needed to explain the 21 cm global spectrum. 

For example requiring $\dot{Q}_{\text {gas }} \sim \Gamma_{C} T_{\mathrm{CMB}}$ at $z=20$ with a $\mathrm{GeV}$ DM particle, one finds that $\sigma^{H} \simeq 10^{-19} \mathrm{~cm}^{2}$. In the case of DM-hydrogen interactions only. It is possible for DM to grow at small relative velocities at most as $v_{\text {rel }}^{-4}$, which is equivalent to a Coulomb-like force, if it is a new fundamental particle obeying the basic rules of relativistic quantum field theory. The following measures are taken to enhance the cross section at low velocities as much as possible. Based on calculations in \cite{Tashiro} and 
\cite{Mu_oz_2015}:

\begin{equation}
\sigma^{I}=\hat{\sigma}^{I} v_{\mathrm{rel}}^{-4} .
\end{equation}

Enhance dark cooling and provide the highest possible cross-sectional area to influence the 21 cm spectrum without violating CMB or direct detection objectives. As $ v_{\mathrm{rel}}=29 \mathrm{~km} / \mathrm{sec} \sim 10^{-4}$ the root mean square relative velocity between the DM and the gas, at decoupling and a later time, it redshifts correspondingly. It can be dismissed if it has a mass well below the mass of a GeV (when sufficiently cooled) and so it doesn't affect gas temperature evolution. The approximate expression can then be determined by assuming a cross section,

\begin{equation}
\dot{Q}_{\text {gas }}^{I} \simeq \sqrt{\frac{2}{\pi}} \frac{\mu_{I}}{m_{I}+m_{p}} \frac{x^{I}}{u_{\text {th }}^{3}}\left(T_{\chi}-T_{\text {gas }}\right) n_{\chi} \hat{\sigma}^{I},
\end{equation}

$\left(u_{\mathrm{th}}^{I}\right)^{2}=T_{\mathrm{gas}} / m_{I}+T_{\chi} / m_{\chi} $ is defined in this section.  Exchanging $\chi \leftrightarrow$ gas yields $\dot{Q}_{\chi}$. Using the above formula, it seems to be possible to obtain a reasonably accurate description of the cooling rate behavior for light dark matter, since the cross section is not dependent of mass.\\

\subsection{21 cm signal and interacting dark energy model}

In cosmology, it is well established that $ 3D $ power spectrum $ P_k $ contains many more modes than the CMB angular power spectrum $ C_l $. Thus, one can use this to discriminate among Dark Energy models. One can also use HI combined signal which is much larger than that of individual sources to have the mapping of large volumes in short periods of time. This technique has already been tested in small surveys \cite{Chang-nature, Santos:2015gra} (references therein), and is a basic method for probing the dark energy models in 21 cm observation.\\

Using the $\Lambda$CDM model to describe the universe, EDGES returns cannot be explained even though $\Lambda$CDM is quite good compared to other observations. As well, this tension is present in the interaction model,\cite{Evidence for interacting dark energy from BOSS} and \cite{Constraints on interacting dark energy models from Planck 2015 and redshift-space distortion data}, we can't find $T_{21}$ as small as the one found in Edges despite using the constraint derived from CMB. For EDGES to meet its data requirements, the interaction should be unreasonably large, which conflicts with previous constraints. However, the interaction reduces the 21 cm value, resulting in a better model than the $\Lambda$CDM model \cite{ed}.

We can use this tension to assert that EDGES may have underestimated the brightness temperature. In fact, the $T_{21}$ should be reduced according to the Wouthuysen-Field effect. Taking the concerns raised in \cite{Concerns about Modelling of the EDGES Data} into account, the EDGES results do need further evaluation. Observations such as these can also constrain dark energy models that have interactions. Hopefully, future experiments will confirm this measurement. This kind of measurement can break the degeneracy between the interaction parameter and the equation of state because, in the standard scenario, brightness temperature is not dependent on the equation of state \cite{ed}.\\

\section{Conclusion}

By interacting with the magnetic moments of the proton and electron, the 21 cm line indicates the hyperfine transition from triplet to singlet of the atomic hydrogen ground state. This splitting leads to two distinct energy levels separated by $\Delta E=5.9 \cdot 10^{-6} \mathrm{eV}$ which corresponds to a photon wavelength $\lambda=2 \pi / \Delta E \simeq 21.1 \mathrm{~cm}$. We reviewed briefly the physics of 21 cm emission in this article.

New experiments such as EDGES, LEDA, and PAPER have enabled probing of the Universe with the 21 cm wavelength. The details of these experiments are described in this review.

Next, the 21 cm emission around the sun was discussed. Radio astronomy observations at wavelengths of 1.42 GHz (equivalent to 21 cm) provide crucial information about the structure and dynamics of the solar atmosphere. It is possible to measure the brightness temperature of the sun, the solar radio flux, and the physical processes that exist in quiet and active regions of the solar atmosphere at this frequency.

A further discussion focused on the role and importance of 21 cm signals on mapping the galaxy and its rotation, and how its measurements and data have led to the discovery of dark matter. Also, a brief description of constraints of 21 cm signal on dark energy was given.

Overall, we can conclude that we are very close to detection of the 21cm signal, close enough to make the move closer to directly observing radio signals absorption or emission from hydrogen atoms to fill gaps in the cosmic timeline. Astrophysics and cosmology are entwined in the 21 cm cosmology. The 21 cm cosmology offers a window into fundamental physics and represents a push on time, sensitivity, and scale frontiers.\\




\begin{thebibliography}{99}

\bibitem{pressbooks}
https://pressbooks.online.ucf.edu/astronomybc/chapter/20-2-interstellar-gas/

\bibitem{lumenlearning}
https://courses.lumenlearning.com/astronomy/chapter/interstellar-gas/

\bibitem{caltech}
https://sites.astro.caltech.edu/~george/reion/


\bibitem{nrao}
https://www.gb.nrao.edu/fgdocs/HI21cm/21cm.html

\bibitem{scienceatyourdoorstep}
https://scienceatyourdoorstep.files.wordpress.com/2018/03/image14.png

\bibitem{hyperphysics}
http://hyperphysics.phy-astr.gsu.edu/hbase/quantum/h21.html

\bibitem{Hyperfine splitting in the ground state of hydrogen}
David J. Griffiths
DOI:10.1119/1.12733

\bibitem{indico}
https://indico.cern.ch/
(Blois DM Jun 4.pdf)


\bibitem{researchgate}
https://www.researchgate.net/figure/The-21cm-cosmological-hydrogen-signal-Top-Time-evolution-of-fluctuations-in-the-21cm


\bibitem{Furlanetto}
Steven Furlanetto (Yale), Michael Furlanetto (LANL),
Astrophysics (astro-ph)
doi:10.1111/j.1365-2966.2006.11169.x
[arXiv:0608067]

\bibitem{Wouthuysen-Field Coupling}
Roy, Ishani and Xu, Wen and Qiu, Jing-Mei and Shu, Chi-Wang and Fang, Li-Zhi,
doi:10.1088/0004-637X/703/2/1992
[arXiv:0908.1695]

\bibitem{global experiments}
http://philosophy-of-cosmology.ox.ac.uk/21cm-background.html

\bibitem{skatelescope}
https://www.skatelescope.org

\bibitem{colorado}
https://www.colorado.edu/ness/projects/experiment-detect-global-eor-signature-edges

\bibitem{Bowman et al}
Judd D. Bowman, Alan E. E. Rogers, Raul A. Monsalve, Thomas J. Mozdzen,  Nivedita Mahesh ,
doi:10.1038/nature25792
[arXiv:1211.3800v2] 


\bibitem{ui.adsabs.harvard.edu}
https://ui.adsabs.harvard.edu/abs/2018MNRAS.478.4193P/abstract

\bibitem{tauceti}
http://www.tauceti.caltech.edu/leda/index.php/science


\bibitem{astro.phy.cam.ac.uk}
https://www.astro.phy.cam.ac.uk/research/research-projects/paper-and-hera


\bibitem{reionization}
https://reionization.org/

\bibitem{arxivhera}
https://arxiv.org/pdf/1606.07473.pdf

\bibitem{berkeley}
http://eor.berkeley.edu/


\bibitem{Probing Radio Intensity}
L. Philip and Z. Abdurashidova and H. C. Chiang and N. Ghazi and A. Gumba and H. M. Heilgendorff and J. Hickish and J. M. Jáuregui-García and K. Malepe and C. D. Nunhokee and J. Peterson and J. L. Sievers and V. Simes and R. Spann,
Instrumentation and Methods for Astrophysics (astro-ph.IM)
[arXiv:1806.09531]

\bibitem{lweb.cfa.harvard.edu}
https://lweb.cfa.harvard.edu/~afialkov
(talk Sievers.pdf)


\bibitem{Patra et al 2013}
Nipanjana Patra, Ravi Subrahmanyan, A. Raghunathan, N. Udaya Shankar,
doi:10.1007/s10686-013-9336-3
[arXiv:1211.3800v2] 

\bibitem{rri}
http://www.rri.res.in/DISTORTION/saras.html

\bibitem{SARAS}
Patra, Nipanjana and Subrahmanyan, Ravi and Sethi, Shiv and Shankar, N. Udaya and Raghunathan, A.
[arXiv:1412.7762]


\bibitem{BIGHORNS}
Sokolowski, Marcin and Tremblay, Steven E. and Wayth, Randall B. and Tingay, Steven J. and Clarke, Nathan and Roberts, Paul and Waterson, Mark and Ekers, Ronald D. and Hall, Peter and Lewis, Morgan and et al.
[arXiv:1501.02922 ]

\bibitem{mwatelescope}
https://www.mwatelescope.org


\bibitem{Voytek et al (2014)}
Tabitha C. Voytek1, Aravind Natarajan, José Miguel Jáuregui García, Jeffrey B. Peterson, and Omar López-Cruz,...
doi:10.1088/2041-8205/782/1/L9
[arXiv:1311.0014 ]


\bibitem{ui.adsabs.harvard.edu2}
https://ui.adsabs.harvard.edu/abs/2015AAS...22531802V/abstract


\bibitem{researchgateSCI-HI}
Tabitha C. Voytek, Aravind Natarajan, Jose Miguel Jauregui-Garcia, Jeffrey B. Peterson, Omar Lopez-Cruz
doi:10.1088/2041-8205/782/1/L9
[arXiv:1311.0014] 

\bibitem{https://www.skatelescope.org/layout/}
https://www.skatelescope.org/layout/

\bibitem{http://www.philbull.com/talks/radio_cosmology_lectures.pdf}
http://www.philbull.com/talks/
(radio cosmology lectures)

\bibitem{Beardsley et al}
Miguel F. Morales, Bryna Hazelton, Ian Sullivan, Adam Beardsley
doi:10.1088/0004-637X/752/2/137
[arXiv:1202.3830]


\bibitem{MWA}
Webster, Rachel,
Proceedings of the International Astronomical Union,
doi=10.1017/S1743921318000893

\bibitem{LOFAR: The LOw-Frequency ARray}
van Haarlem, Michael P., et al. "LOFAR: The low-frequency array." Astronomy and astrophysics 556 (2013): A2.
https://doi.org/10.1051/0004-6361/201220873


\bibitem{http://www.lofar.org/}
http://www.lofar.org/

\bibitem{http://www.astronomynotes.com/ismnotes/s3.htm}
http://www.astronomynotes.com/ismnotes/s3.htm

\bibitem{21 cm Intensity Mapping}
Jeffrey B. Peterson, Roy Aleksan, Reza Ansari, Kevin Bandura,...
[arXiv:0902.3091]

\bibitem{https://ned.ipac.caltech.edu/level5/Sept10/Turner/Turner6.html}
https://ned.ipac.caltech.edu/level5/Sept10/Turner/Turner6.html

\bibitem{BAO}
Chang, Tzu-Ching and Pen, Ue-Li and Peterson, Jeffrey B. and McDonald, Patrick,...
Phys.Rev.Lett.100:091303,2008
doi:10.1103/PhysRevLett.100.091303
[arXiv:0709.3672 ]

\bibitem{intensity mapping}
Jeffrey B. Peterson and Roy Aleksan and Reza Ansari ,...
Instrumentation and Methods for Astrophysics (astro-ph.IM)
[arXiv:0902.3091 ]


\bibitem{http://joans.io/papers/21cm.pdf}
http://joans.io/papers/21cm.pdf


\bibitem{Pawsey and Yabsley 1949}
Pawsey, J. L., and D. E. Yabsley. "Solar radio-frequency radiation of thermal origin." Australian Journal of Chemistry 2.2 (1949): 198-213.

\bibitem{Denisse 1949}
Microwave solar noise and sunspot,Denisse, J. F. 
doi: 10.1086/106280


\bibitem{Krishnan Labrum}
Krishnan, T. ; Labrum, N. R.
doi:10.1071/PH610403

\bibitem{Waldmeier and Muller 1950}
Waldmeier and Muller 1950,
Zeitschrift fur Astrophysik, Vol. 27, p.58

\bibitem{Christiansen Warburton Davies}
Christiansen, W. N. ; Warburton, J. A. ; Davies, R. D.
doi: 10.1071/PH570491 


\bibitem{Christiansen and Mathewson (1958)}
CHRISTIANSEN, W. N. and MATHEWSON, D. S.: 1958, Proc. Inst. Radio Engr. 46 ,127. 


\bibitem{Christiansen and Warburton (1955)}
The Sun in two dimensions at 21 CM
Christiansen, W. N. ; Warburton, J. A.
The Observatory, Vol. 75, p. 9-10 (1955)

\bibitem{Christiansen and Warburton (1955)2}
The Distribution of Radio Brightness over the Solar Disk at a Wavelength of 21 Centimetres. III. The Quiet Sun-Two-Dimensional Observations,Christiansen, W. N. ; Warburton, J. A.


\bibitem{http://adsabs.harvard.edu/full/1961AuJPh..14..403K}
http://adsabs.harvard.edu/full/1961AuJPh..14..403K


\bibitem{A study of a solar active region using combined optical and radio techniques}
Christiansen, W. N. ; Mathewson, D. S. ; Pawsey, J. L. ; Smerd, S. F. ; Boischot, A. ; Denisse, J. F. ; Simon, P. ; Kakinuma, T. ; Dodson-Prince, H. ; Firor, J.
Bibcode: 1960AnAp...23...75C 


\bibitem{constraining dark matter}
Valdés, M. and Ferrara, Andrea and Mapelli, Michela and Ripamonti, Emanuele,
doi =10.1111/j.1365-2966.2007.11594.x

\bibitem{https://astronomynow.com/2016/11/08/new-theory-of-gravity-might-explain-dark-matter/}
https://astronomynow.com 
(new theory of gravity might explain dark-matter)


\bibitem{kolb}
Edward W. Kolb and Michael S. Turner. Addison-Wesley, Redwood City, CA,
doi:10.1126/science.249.4970.808-a

\bibitem{Diemand Moore 2011}
J. Diemand, M. Kuhlen, P. Madau, M. Zemp, B. Moore, D. Potter, J. Stadel,
doi:10.1038/nature07153
[arXiv:0805.1244 ]

\bibitem{Frenk White 2012}
Carlos S. Frenk, Simon D. M. White,
doi:10.1002/andp.201200212
[arXiv:1210.0544v1]

\bibitem{Signs of Dark Matter at 21-cm}
Barkana, Rennan and Outmezguine, Nadav Joseph and Redigolo, Diego and Volansky, Tomer,
Phys. Rev. D 98, 103005 (2018)
doi:10.1103/PhysRevD.98.103005
[arXiv:1803.03091]


\bibitem{Bode et al 2001}
Bode, Paul ; Ostriker, Jeremiah P. ; Turok, Neil,
doi:10.1086/321541 
[arXiv:astro-ph/0010389]

\bibitem{Schneider et al 2012}
Caroline A Schneider, Wayne S Rasband ,Kevin W Eliceiri,
doi:10.1038/nmeth.2089

\bibitem{Viel et al 2013}
M. Viel, G.D. Becker, J.S. Bolton, M.G. Haehnelt,
doi:10.1103/PhysRevD.88.043502
[arXiv:1306.2314] 

\bibitem{Tashiro}
Tashiro, Hiroyuki and Kadota, Kenji and Silk, Joseph,
doi:10.1103/PhysRevD.90.083522
[arXiv:1408.2571 ]


\bibitem{Mu_oz_2015}
Muñoz, Julian B. and Kovetz, Ely D. and Ali-Haïmoud, Yacine,Phys. Rev. D 92, 083528 (2015)
doi:10.1103/PhysRevD.92.083528
[arXiv:1509.00029]


\bibitem{Barkana}
Barkana, Rennan,
Nature 555 (2018) 71-74
doi:10.1038/nature25791
[arXiv:1803.06698 ]

\bibitem{Chang-nature}
Hydrogen 21-cm Intensity Mapping at redshift 0.8
Tzu-Ching Chang, Ue-Li Pen, Kevin Bandura, Jeffrey B. Peterson
[arXiv:1007.3709 [astro-ph.CO]]
Nature 466.7305 (2010): 463-465.
%
\bibitem{Santos:2015gra}
M.~G.~Santos, P.~Bull, D.~Alonso, S.~Camera, P.~G.~Ferreira, G.~Bernardi, R.~Maartens, M.~Viel, F.~Villaescusa-Navarro and F.~B.~Abdalla, \textit{et al.}
PoS \textbf{AASKA14}, 019 (2015)
doi:10.22323/1.215.0019
[arXiv:1501.03989 [astro-ph.CO]].

\bibitem{ed}
Andre A. Costa, Ricardo C. G. Landim, Bin Wang, E. Abdalla
doi:10.1140/epjc/s10052-018-6237-7
[arXiv:1803.06944]






\bibitem{Evidence for interacting dark energy from BOSS}
Elisa G. M. Ferreira, Jerome Quintin, André A. Costa, E. Abdalla, Bin Wang
doi:10.1103/PhysRevD.95.043520
[arXiv:1412.2777] [astro-ph.CO]

\bibitem{Constraints on interacting dark energy models from Planck 2015 and redshift-space distortion data}
André A. Costa, Xiao-Dong Xu, Bin Wang, E. Abdalla
doi:10.1088/1475-7516/2017/01/028
[arXiv:1605.04138] [astro-ph.CO]

\bibitem{Concerns about Modelling of the EDGES Data}
Richard Hills, Girish Kulkarni, P. Daniel Meerburg, Ewald Puchwein
doi:10.1038/s41586-018-0796-5
[arXiv:1805.01421 ][astro-ph.CO]


\bibitem{modifiedgravity}
T.~Clifton, P.~G.~Ferreira, A.~Padilla and C.~Skordis,
Phys. Rept. \textbf{513} (2012), 1-189
doi:10.1016/j.physrep.2012.01.001;
S.~Nojiri, S.~D.~Odintsov and V.~K.~Oikonomou,
Phys. Rept. \textbf{692}, 1-104 (2017)
doi:10.1016/j.physrep.2017.06.001


\end{thebibliography}
\end{document}